# Signatures of Weak Spin-Phonon Coupling in 12R-BaTi$_{0.4}$Mn$_{0.6}$O$_3$: A Raman Study

Bommareddy Poojitha, Anjali Rathore, and Surajit Saha[a)]

*Indian Institute of Science Education and Research, Bhopal, Bhauri campus 462066, INDIA*
[a)]*Corresponding Author: surajit@iiserb.ac.in*

**Abstract.** Antiferromagnetic materials with strong spin-phonon coupling have the potential for spintronic applications. Here, we report the structural, magnetic, and vibrational properties of 12R-BaTi$_{0.4}$Mn$_{0.6}$O$_3$. Temperature-dependent magnetization data show the presence of weak antiferromagnetic interactions. Correlation between magnetic and vibrational properties are probed by using temperature-dependent Raman spectroscopic measurements. Two of the Raman active phonons show deviation from anharmonic behaviour at low temperatures which can be attributed to spin-phonon coupling. The strength of coupling is estimated using mean-field approximation and is found to be 1.1 and 1.2 cm$^{-1}$ for phonons at 530 and 717 cm$^{-1}$, respectively.

## INTRODUCTION

Antiferromagnets with rich spin-dynamics and spin-phonon coupling are of great interest in the emerging field of THz magnonics and spintronics [1]. Multiferroics are the materials which exhibit more than one primary ferroic order parameters simultaneously. BaTiO$_3$ is well-known room temperature ferroelectric compound. New multiferroic materials can be contrived by introducing magnetic ions such as Mn, Fe, Co, Cr etc. into conventional ferroelectric materials, like BaTiO$_3$. Hiroyuki NAKAYAMA et. al. [2] have performed ab-initio total-energy calculations and reported that Mn is the most promising candidate to dope BaTiO$_3$ among all magnetic ions such as Sc, V, Cr, Mn, Fe, Co, Ni, Cu. BaTi$_{1-x}$Mn$_x$O$_3$ system is predicted to be a good material to induce multiferroicity. On the other hand, it was experimentally observed that BaTiO$_3$ undergoes structural transition upon chemical doping. For example, Eu$^{+2}$ doping induces tetragonal to cubic transition [3], and the partial substitution of Mn at Ti site in tetragonal BaTiO$_3$ leads to a series of structural transitions which make this system to be a good platform to explore different crystal phases and study the structure-property relationship [4]. The ferroelectricity intimately depends on the structure of the unit cell, thus showing a serious and immediate consequence of any distortion. The crystal structure adopted by perovskite compounds can be estimated using the measure of Goldsmith tolerance factor which can be written as $t = \frac{r_A + r_O}{\sqrt{2}((1-x)r_B + xr_{B'} + r_O)}$ for B-site doped manganite compounds with the chemical formula AB$_{1-x}$B'$_x$O$_3$. t ~ 1 leads the system to adopt an ideal cubic structure where all BO$_6$ octahedra are corner shared and B-O-B bond angle is nearly 180º. t >> 1 leads to face shared octahedra where B-O-B bond angle is nearly 90º. t > 1 results in the intermediate structure where the combination of corner-sharing (c-cubic) as well as face-sharing (h-hexagonal) octahedra are present. For example, tetragonal to hexagonal (6H) transition in BaTiO$_3$ is extensively studied in the literature, 6H unit cell consists of (cch)$_2$ stacking sequence. In Mn-rich compositions of BaTi$_{1-x}$Mn$_x$O$_3$ system, 12R, 10H, 9R, 8H, and 15R phases are observed in [5]. Among all polymorphs, 12R phase has (hhcc)$_3$ stacking sequence with an equal proportion of corner and face sharing connectivity. 12R is also known to be a trademark of the frustrated system due to its crystal structure and also magnetic properties [6]. It has low symmetry and a high degree of chemical disorder in terms of site occupancy. It is interesting to verify and investigate the correlation between these frustrated spins and phonons.

In this article, we report the correlation between magnetism and vibrational properties of 12R-BaTi$_{0.4}$Mn$_{0.6}$O$_3$. The crystal structure is examined by powder x-ray diffraction. Temperature-dependent Raman studies reveal the presence of weak spin-phonon coupling. The strength of coupling is estimated using mean-field approximation.

## EXPERIMENTAL DETAIL

Polycrystalline sample, $BaTi_{0.4}Mn_{0.6}O_3$ (BTM60) was prepared by a conventional high-temperature solid-state reaction method. High purity powders of $BaCO_3$, $TiO_2$, and $MnO_2$ were used as precursor materials. The stoichiometric mixture was calcined twice and sintered once at 1250°C, 1300°C, and 1400°C, respectively for 2 hours each with intermediate grindings. X-ray reflections were collected using PANalytical Empyrean X-ray diffractometer with Cu-$K_\alpha$ radiation of wavelength 1.5406 Å. Raman spectra were recorded on $BaTi_{0.4}Mn_{0.6}O_3$ pellet using a LabRAM HR Evolution Raman spectrometer attached with Nd: YAG laser of 532 nm wavelength and Peltier cooled charge-coupled device (CCD) detector. HFS600E-PB4 Linkam stage was used to control and monitor temperature during Raman measurements. DC magnetization measurements were done by using Quantum Design SQUID-VSM (Superconducting Quantum Interference Device with Vibrating Sample Magnetometer).

## RESULTS AND DISCUSSION

X-ray reflection patterns of BTM60 collected at room temperature are shown in Figure 1. Rietveld refinement results reveal that the BTM60 is stabilized in 12R-type hexagonal structure with the space group R-3m (No. 166). The unit cell is drawn using VESTA (Visualization for Electronic and Structural Analysis) software and displayed as an inset of Figure 1(b). The unit cell consists of $Mn_3O_{12}$ trimers (formed by three face-shared octahedra) connected to each other via corner-shared octahedron. According to the conclusion drawn from the refinement of x-ray and neutron diffraction data by G. M. Keith et al., [8] Mn atoms have strong tendency to occupy the central site of face-sharing trimers (M1), and Ti atoms have a strong preference for outer sites of face-sharing trimers (M2) and corner-sharing octahedra (M3).

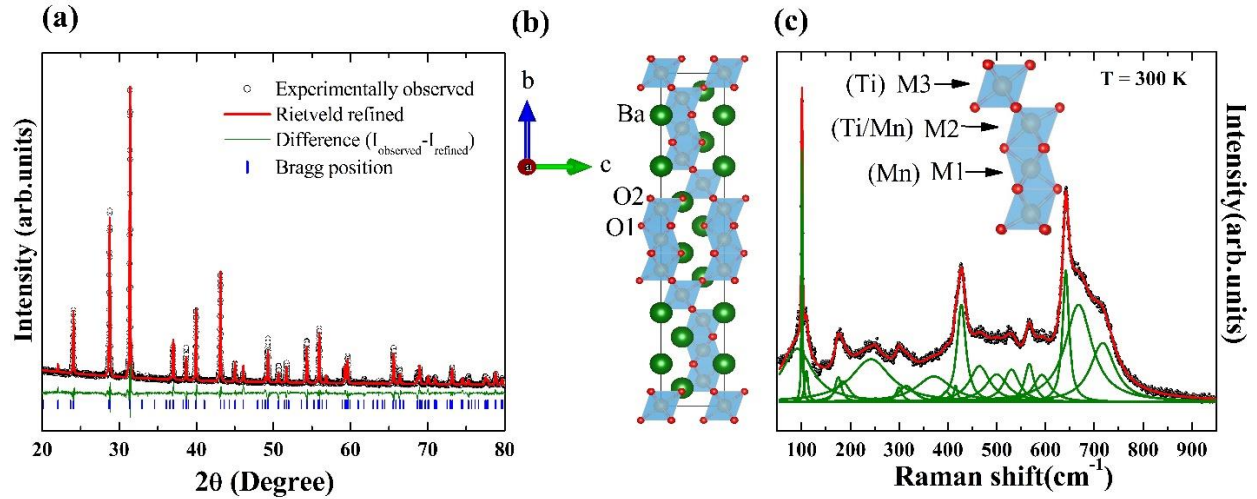

**FIGURE 1**. (a) X-ray diffraction pattern, (b) Unit cell, and (c) Raman spectra of $BaTi_{0.4}Mn_{0.6}O_3$ at room temperature (symbols are experimental data and solid lines in (c) represent the Lorentzian fit).

The refined lattice parameters at room temperature are a = 5.6842 Å and c = 27.882 Å which are falling in line with reports on similar compositions [5,7,8]. Figure 1(c) shows the Raman spectrum at room temperature. Group theory predicts 24 Raman active modes for 12R phase with irreducible representations $\Gamma_{Raman} = 9A_{1g} + 15E_g$ where both M1 and M3 refrain from participating in the Raman modes [9]. Nineteen Raman modes could be resolved at room temperature and their frequencies are 101, 175, 185, 242, 252, 300, 315, 371, 396, 416, 428, 465, 500, 530, 567, 593, 668, 642, and 717 $cm^{-1}$. To investigate the phonon behaviour for observed phonons, we have performed Raman measurements in the temperature range of 80 to 450 K. The frequency for most of the phonons is observed to be decreasing with increasing temperature which can be explained by Balkanski's anharmonic model [10] (for example, the mode at 642 $cm^{-1}$ in Figure 2(b)). However, the modes at 101, 530, and 717 $cm^{-1}$ show deviation from anharmonic behaviour at low temperatures (will be referred to as anomaly in later discussions). In general, the frequency of Raman active phonon as a function of temperature can be written as [10,11]:

$$\omega(T) = \Delta\omega_{anh}(T) + \Delta\omega_{el-ph}(T) + \Delta\omega_{sp-ph}(T)$$

here, $\Delta\omega_{anh}(T)$ accounts for the change in phonon frequency due to intrinsic anharmonicity while $\Delta\omega_{el-ph}(T)$ corresponds to the change due to electron-phonon interactions which is absent in this system and $\Delta\omega_{sp-ph}(T)$ accounts for spin-phonon coupling. The shift in the phonon frequency due to cubic anharmonicity can be expressed as [10,11]:

$$\omega_{anh}(T) = \omega_0 + A\left[1 + \frac{2}{\left(e^{\frac{\hbar\omega_0}{2k_B T}} - 1\right)}\right] \quad (1)$$

where, $\omega_0$ is the frequency of the phonon at absolute zero temperature, $A$ is cubic anharmonic coefficient for frequency, $\hbar$ is reduced Planck constant, $k_B$ is Boltzmann constant and $T$ is the variable temperature.

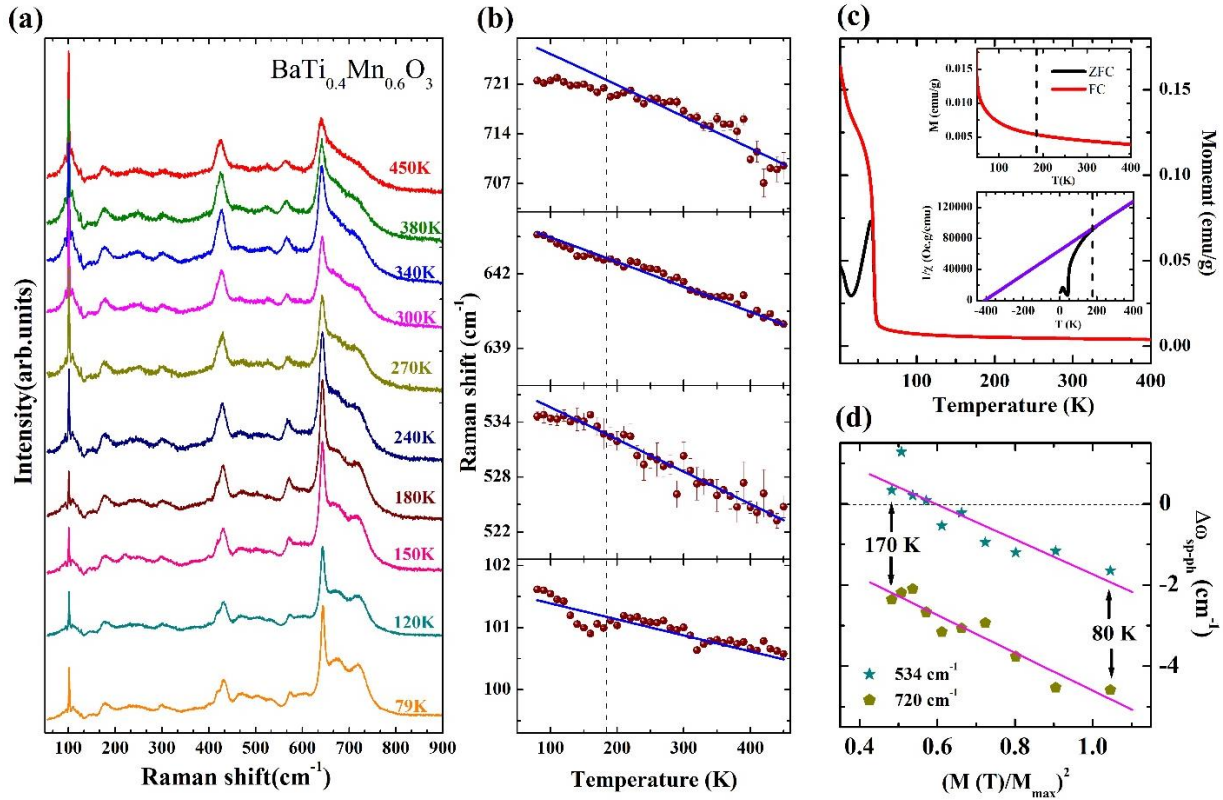

**FIGURE 2**. (a) Raman spectra at a few temperatures, (b) phonon frequency as a function of temperature, (c) temperature-dependent magnetization, insets of (c) are enlarged view of M(T) in high-temperature region and inverse susceptibility as a function of temperature, (d) plot of $\Delta\omega_{sp-ph}$ vs $\left(\frac{M(T)}{M_{max}}\right)^2$ (see text).

Unlike the mode at 642 cm$^{-1}$, the modes at 530 and 717 cm$^{-1}$ show an anomalous deviation from the anticipated anharmonic behavior (shown by solid lines in Figure 2b). In order to probe the possible origin of the deviation we have performed temperature-dependent magnetization measurements. Since 101 cm$^{-1}$ mode is associated with Ba vibrations, the fluctuations in frequency below 170 K may be due to an indirect effect of magnetic order on lattice.

Magnetization with varying temperature was measured at both zero-field cooling (ZFC) and field cooling (FC) under an external field of 500 Oe between 2 - 400 K. It is very clear from Figure 2 (c) that the ZFC and FC curves overlap each other down to 42 K while a clear bifurcation happens at lower temperatures. This bifurcation between ZFC and FC may be due to a possible ferrimagnetic ordering at low temperatures owing to multi-valency of Mn ions (Mn$^{+4}$ and

$Mn^{+3}$) which is, therefore, not an intrinsic property of BTM60 [7,8]. However, temperature-dependent inverse susceptibility fits well with Curie-Weiss equation above 170 K. Though there is no clear signature of long-ranged antiferromagnetic ordering, the Weiss temperature is strongly negative (Inset of Figure 2(c)) which indicates a possible presence of short-ranged antiferromagnetic interaction in the system [5,7,8]. The interaction is short-ranged because the Ti atoms are located in the corner-shared octahedra while the Mn spins are at the center of the trimers (face-shared octahedra) thus making Mn-Mn distance effectively 9.8543 Å. Such long distances between Mn spins prevent any long-range magnetic order down to 2K.

**Signatures of spin-phonon coupling:** In magnetic materials, in presence of spin-spin correlations a phonon mode may get renormalized due to spin-phonon interactions leading to phonon anomalies (like the modes at 530 and 717 cm$^{-1}$). We have used mean-field approximation to estimate the spin-phonon coupling constant for the anomalous modes. Under this approximation, the temperature dependence of phonon can be expressed as [12,13]; $\omega(T) = \omega_{anh}(T) + \lambda <S_i.S_j>$ where, $\omega_{anh}$ is the phonon frequency in the absence of spin-spin interactions, $<S_i.S_j>$ is the spin-spin correlation function, and $\lambda$ is the strength of spin-phonon coupling. Under the mean-field approximation, the above equation gets the form; $\lambda <S_i.S_j> \approx n\lambda \left(\frac{M(T)}{M_{max}}\right)^2$. Here n is the in-plane coordination number for magnetic ion while M(T) and M$_{max}$ are temperature-dependent magnetization and the maximum magnetization, respectively. The renormalization of phonon frequency due to spin-phonon coupling below 170 K can be written as; $\Delta\omega_{sp-ph} = \omega(T) - \omega_{anh}(T) = \lambda <S_i.S_j>$. Figure 2(d) shows the plot of $\Delta\omega_{sp-ph}$ vs $n\lambda \left(\frac{M(T)}{M_{max}}\right)^2$ for the phonons at 530 and 717 cm$^{-1}$ showing a linear trend thus confirming the presence of spin-phonon coupling. The coupling constant ($\lambda$) is found to be 1.1, and 1.2 cm$^{-1}$ for 530, and 717 cm$^{-1}$ modes, respectively, which represents weak coupling [14-17].

## CONCLUSION

We have synthesized polycrystalline sample of 12R-BaTi$_{0.4}$Mn$_{0.6}$O$_3$ using conventional solid-state reaction route. Two phonon modes are observed to be anomalous at low temperatures due to a weak spin-phonon coupling arising from short-ranged antiferromagnetic ordering below 170 K. The coupling strengths are estimated by using mean-field approximation.

## ACKNOWLEDGEMENTS

Funding from DST/SERB (project No. ECR/2016/001376), Nanomission (Project No. SR/NM/NS-84/2016(C)) and DST-FIST (Project No. SR/FST/PSI-195/2014(C)) are acknowledged. B.P acknowledges UGC India for fellowship.